\documentclass[12pt,preprint,aps,english,superscriptaddress]{revtex4-1}
\usepackage[T1]{fontenc}
\usepackage[latin9]{inputenc}
\usepackage{amsmath}
\usepackage{amssymb}
\usepackage{cancel}
\usepackage{graphicx}
\usepackage{esint}
\usepackage{color}
\makeatletter

\usepackage{epstopdf}

\@ifundefined{textcolor}{}{%
 \definecolor{BLACK}{gray}{0}
 \definecolor{WHITE}{gray}{1}
 \definecolor{RED}{rgb}{1,0,0}
 \definecolor{GREEN}{rgb}{0,1,0}
 \definecolor{BLUE}{rgb}{0,0,1}
 \definecolor{CYAN}{cmyk}{1,0,0,0}
 \definecolor{MAGENTA}{cmyk}{0,1,0,0}
 \definecolor{YELLOW}{cmyk}{0,0,1,0}
}

\usepackage{dcolumn}
\usepackage{bm}
\usepackage{enumerate}

\usepackage[caption=false]{subfig}

\newcommand{\captiontitle}[1]{{\bf #1}}

\@ifundefined{showcaptionsetup}{}{%
\PassOptionsToPackage{caption=false}{subfig}}

\usepackage{babel}
\addto\captionsenglish{}
\renewcommand{\thefigure}{\@arabic\c@figure}

\makeatother

\begin{document}
\title{Multi-flat bands and strong correlations in Twisted Bilayer Boron Nitride}

\author{Lede Xian}
\email{These authors contributed equally}
\affiliation{Max Planck Institute for the Structure and Dynamics of Matter, Luruper Chaussee 149, 22761 Hamburg, Germany}

\author{Dante M. Kennes}
\email{These authors contributed equally}
\affiliation{Dahlem Center for Complex Quantum Systems and Fachbereich Physik,
Freie Universit\"{a}t Berlin, 14195 Berlin, Germany}

\author{Nicolas Tancogne-Dejean}
\affiliation{Max Planck Institute for the Structure and Dynamics of Matter, Luruper Chaussee 149, 22761 Hamburg, Germany}

\author{Massimo Altarelli}
\affiliation{Max Planck Institute for the Structure and Dynamics of Matter, Luruper Chaussee 149, 22761 Hamburg, Germany}

\author{Angel Rubio}
\email{angel.rubio@mpsd.mpg.de}
\affiliation{Max Planck Institute for the Structure and Dynamics of Matter, Luruper Chaussee 149, 22761 Hamburg, Germany}
\affiliation{Center for Computational Quantum Physics (CCQ), The Flatiron Institute, 162 Fifth Avenue, New York, NY 10010, USA}

\date{\today}

\maketitle

{\bf Effectively two-dimensional materials rank among the most fascinating systems of contemporary condensed matter research and beyond. They can exhibit interesting phases of matter, such as the quantum hall state \cite{Stern08}, are vital in quantum computing, due to their unique braiding properties and can host anyons \cite{ivanov00,nayak07}, which fundamentally refines our classification of particles in fermions and bosons. They, furthermore, still pose a grand challenge to theoretical tools; which lack a flagship method describing two-dimensional correlated systems. This has, e.g., given rise to the field of quantum simulations of two-dimensional systems in cold gases \cite{Kondov15,Gross17}. In a groundbreaking experimental advance \cite{cao2018a,cao2018b,Yankowitz18} it was recently shown that by stacking two sheets of graphene atop of each other at a twist angle close to one of the so called ``magic angles'' \cite{bistritzer2011,Tarnopolsky18}, an effective two-dimensional correlated system emerges. In this system the kinetic energy of the low-energy electrons is much reduced and consequently interactions become very relevant, providing a new platform into the physics of two-dimensional correlated materials. Evidence of a proposed Mott insulating as well as superconducting state in these highly tunable systems has spurred much attention \cite{cao2018a,cao2018b} as they could pave the way to understanding long-standing questions of high-$T_c$ superconductivity \cite{Anderson2013} or provide candidate systems for topological chiral superconductors; key to highly relevant  quantum technologies \cite{Stern08}. Here, we demonstrate that twisted bilayer boron nitride (TBBN) is an exciting and even richer alternative to twisted bilayer graphene (TBG). Crucially, we show that in TBBN multiple flat bands emerge without having to fine tuning close to a ``magic angle'' that upon doping lead to correlated phases of matter (insulating and superconducting). TBBN  could thus be much less sensitive to small deviations in the twist angle and therefore provide a particularly suited experimental platform to study correlation physics in two dimensions. Furthermore, we find that in marked contrast to TBG at small twist angle families of 2,4 and 6-fold degenerate, well separated, bands emerge within the gap, considerably broadening the addressable physics.   }

Graphene is considered a very special 2D material, as its symmetries give rise to gapless linear dispersions for the low-energy bands around the Fermi level at the corner of the Brillouin zone \cite{Morell10,Kim17}. These low energy states can be well described by a massless Dirac Hamiltonian, which gives rise to many of graphene's unusual properties, such as lack of back-scattering and Klein tunneling \cite{neto2009}. So far, it is unclear whether this special gapless and massless features are crucial in the observed strongly-correlated-like phenomena \cite{cao2018a,cao2018b,Yankowitz18} in twisted bilayer systems \cite{Wu17}. Here, we elaborate on this question by studying a similar system,  twisted bilayer boron nitride (TBBN), which however has a large band gap. Hexagonal boron nitride (hBN) is also called ``white graphene'', because it has similar hexagonal structure as graphene \cite{Wirtz06,Blase95}. While graphene has two identical carbon atoms in the unit cell, they are replaced by one nitrogen and one boron atoms in hBN. This difference gives rise to a large band gap and finite effective mass for the low-energy electrons in hBN. As a consequence, hBN is widely used as an ideal insulator in 2D electronic devices (and to enhance the properties of graphene by isolating it from its environment). Interestingly, our density functional theory (DFT) calculations reveal that flat bands develop at both the top of the valence bands and bottom of the conduction bands in twisted bilayer hBN with a small twist angle. In marked contrast to the TBG systems, the band width of these flat bands decrease monotonically with twist angle and there is no occurrence of  ``magic angles'' \cite{bistritzer2011}. We further evaluate the Coulomb interaction for those localized states in the flat bands using a self-consistent \textit{ab inito} DFT+$U$ \cite{Nicolas17} and functional renormalization group (FRG) \cite{Metzner2012a} method to show that the flat band at the top of the valence bands can host exotic strongly-correlated physics, such as the appearance of a Mott insulator phase and unconventional superconductivity for slight amount of doping or photo-doping.  Those results provide strong evidence that TBBN could also be utilized to study unconventional superconductors, much like TBG. Our results suggest that the gapless-ness and massless-ness  of graphene is not essential in the observed exotic strongly-correlated phenomena motivating studies of many more two-dimensional materials stacked with a small twist angle.

\begin{figure}[t]
\centering
\includegraphics[width=\columnwidth]{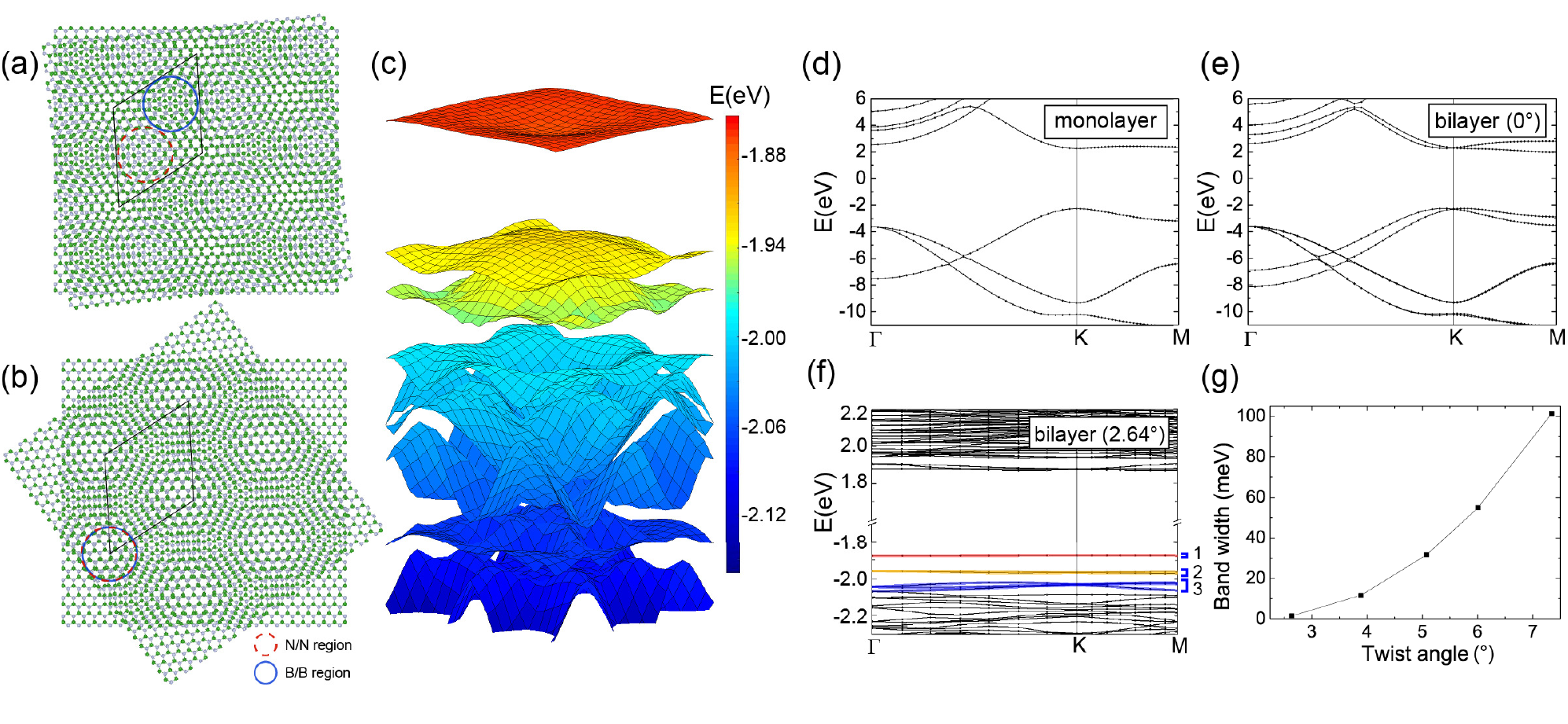}
\caption{\captiontitle{Atomic and electronic structures of TBBN} (a) and (b) Schematic illustration of the two possible configurations in twisted bilayer hBN that have the same Moir\'{e} pattern. The B/B regions (highlighted with solid blue circle) and the N/N regions (highlighted with dash red circle) are located in different sites in the supercell in configuration {$\alpha$} (a), while they share the same sites in configuration {$\beta$} (b). (c) DFT band structure of unrelaxed twisted bilayer hBN at $2.64^\circ$ in a region of $0.11  \times 0.11$ $1/\text{\normalfont\AA}^2$ around the $\Gamma$ point in the supercell Brillouin zone. (d-f) DFT band structures of monolayer (d), normal bilayer without twist (e) and unrelaxed twisted bilayer hBN (f). (g) Band width of the first set of flat bands at the top of the valence bands of unrelaxed twisted bilayer graphene for different twist angles. }
\label{fig:fig1}
\end{figure}

Starting from a regular bilayer hBN with stacking as in the bulk,  TBBN can be formed by twisting the top or the bottom layer with angle $\theta$. As shown in Fig.~\ref{fig:fig1}(a) and (b), there are two possible, distinct configurations of TBBN, called, $\alpha$ and $\beta$, that lead to the same supercell. Their twist angles differ by $60^\circ$. In TBBN, there are regions where nitrogen atoms on the top layer are approximately on top of those of the bottom layer (the N/N region, as highlighted by dash red circles) and regions where the same holds for the boron atoms (the B/B region, as highlighted by solid blue circles). In configuration $\alpha$, the two regions are located at different sites of the supercell, while in configuration $\beta$, they are located at the same sites (see Fig.~\ref{fig:fig1}(a) and (b)).

To study the effect of twisted stacking, we first investigate the band structures of TBBN with fixed interlayer separation $d$, which is set to be the optimized interlayer distance of bilayer hBN without twist (3.23 Angstrom, that is very close to the experimental value). As shown in Fig.~\ref{fig:fig1}(c) and (f), similar to the case of TBG, flat bands appear at the top of the valence band in TBBN at small twist angles. These flat band structures are significantly different from those of monolayer and normal bilayer hBN shown in Fig.~\ref{fig:fig1}(d) and (e), respectively. In contrast to the case of graphene, the flat bands are isolated from the other bands with a relatively large gap of over 100meV, and intriguingly, more than one set of them develop as twist angle decreases. For the smallest twist angle we studied in (Fig.~\ref{fig:fig1}(f)), three set of flat bands can be clearly identified (see the labels on the right side of Fig.~\ref{fig:fig1}(f)) and the number of bands in each set is 2, 4 and 6. The band width of these flat bands decreases significantly  with the twist angle. As shown in Fig.~\ref{fig:fig1}(g), the band width of the top of the valence bands (the first set of flat bands) decreases from 101.4 meV at $7.34^\circ$ to 1.6 meV at $2.64^\circ$. Modeling the top of the valence bands of hBN by a massive Dirac Hamiltonian suggests that indeed the band width of these flat bands will decrease monotonically with twist angles without the existence of any ``magic angles'' (see supplemental materials). Similarly, flat bands  also appear at the bottom of the conduction bands as the twist angle decreases (see Fig.~\ref{fig:fig1}(f) at around $1.87\sim1.90$ eV). 

\begin{figure}[t]
\centering
\includegraphics[width=\columnwidth]{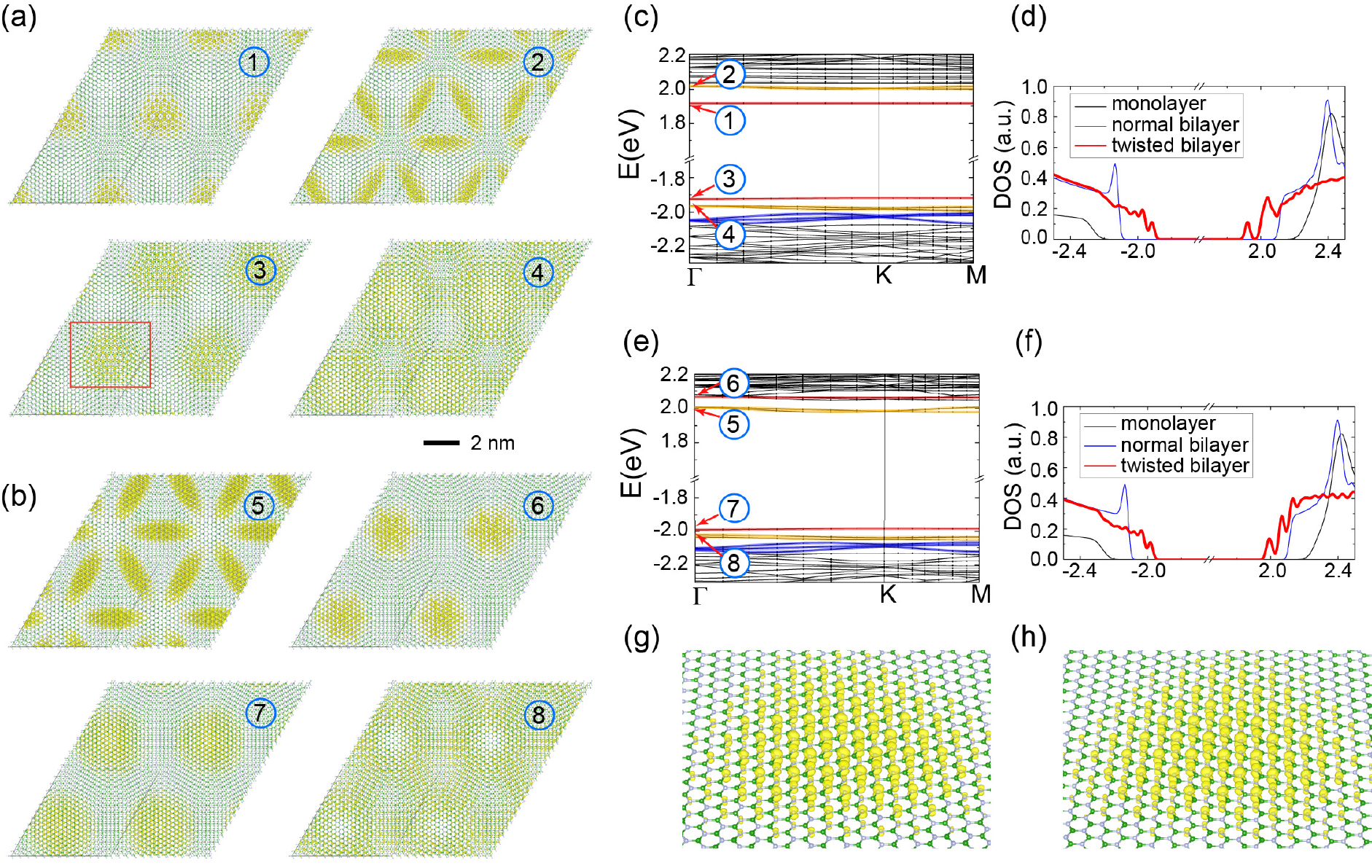}
\caption{\captiontitle{Localization of electronic states in the flat bands of TBBN at $\mathbf{2.64^\circ}$ } (a) and (b) Top view of the charge density plot for the electronic states highlighted in the band structures shown in (c) and (e) for configurations {$\alpha$} and {$\beta$} after relaxation, respectively. (d) and (f) The corresponding density of states of twisted bilayer hBN in the two configurations in comparison to those of monolayer and bilayer hBN without twist.  (g) and (h) Prospective view of the charge density localization region highlighted in red in (a) for the bottom (g) and top (h) layers. }
\label{fig:fig2}
\end{figure}

The calculated DFT band structures for the two configurations $\alpha$ and $\beta$ of TBBN with {\it fixed} interlayer separation are almost identical, showing only subtle differences, while they are noticeably different {\it after relaxation}. To further understand the effect of the structural relaxation in the two configurations and the nature of the flat bands, we  plot the charge density distribution for the low-energy states in the two configurations of TBBN in Fig.~\ref{fig:fig2}. Consistent with their narrow band width, those low-energy states are well localized in real space, with different characters in different flat bands. As expected, the low-energy states at the top of the valence bands are dominated by nitrogen $p_z$ orbital and those at the bottom of the conduction bands are dominated by boron $p_z$ orbital (see Fig.~\ref{fig:fig2}(g) and (f)). Therefore, the flat-band states of the valence bands are localized around the N/N region, while those of the conduction bands are localized around the B/B region. The relaxation increases the interlayer separation in the N/N region for both configurations and the valence bands of the two configurations remain very similar. However, we found that the response to the relaxation in the B/B regions is different for the two configurations, because of their different relative position with respect to the N/N region. This difference leads to different energy ordering of these flat band states in the conduction bands for the two configurations. Along with the different response to the relaxation, the band gaps of the two configurations are also different. In the total density of states, shown in Fig.~\ref{fig:fig2}(d) and (f), the low energy flat bands manifest as a few sharp peaks near the band edges, which are significantly different from those of the monolayer and untwisted bilayer hBN.

Next, in order to characterize the electronic correlations in those flat bands (that tend to be beyond the capabilities of standard local and semi-local XC functionals) we investigate TBBN by means of self-consistent DFT+$U$ (see Methods). We aim at extracting the most \textit{ab initio} information possible concerning the electron-electron correlations in TBBN, that afterwards  will be  used to motivate the functional renormalization group (FRG) treatment (see below).
We focus on the bands labeled 3 in Fig.~\ref{fig:fig2} hereafter. These are two-fold degenerated bands, well described by a two-band Hubbard model. Concentrating on local interactions in such a model, one needs to determine an on-site Hubbard $U$ in addition to the Hunds-coupling $J$. Assessing how important $J$ is, from \textit{ab initio} calculations, is crucial to our model building of TBBN.

\begin{figure}[t]
\centering
\includegraphics[width=\columnwidth]{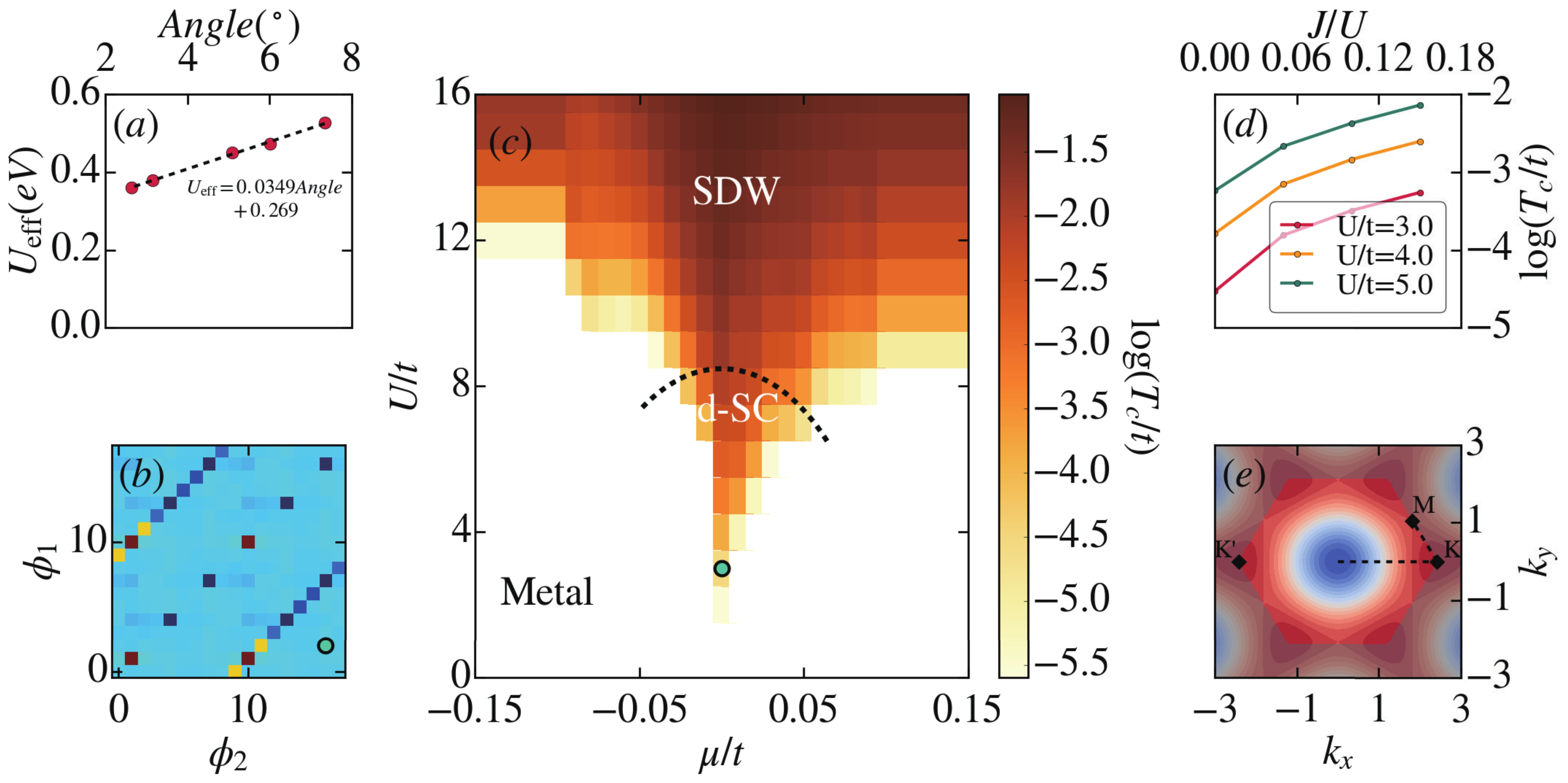}
\caption{\captiontitle{Correlations in Twisted Bilayer Boron Nitride} (a) interaction parameter as obtained by DFT+$U$ without doping. (b)  Structure of the effective two-particle interaction $\Gamma(k_1,k_2,k_3)$ at the end of the FRG flow for fixed $\phi_3$ (angle of momentum $k_3$) in dependence of the remaining two angles $\phi_1$ and $\phi_2$ (of momenta $k_1$ and $k_3$) for $U=3t $ as well as $\mu/t=0$ and $J=0$. Dominant features along the diagonal (incoming momenta sum to zero)  as well as the alternating sign (negative (red) to positive (blue)) signal a d-wave pairing instability. (c) Phase diagram in dependence of $U/t$ and $\mu/t$ for $J=0$ (values of $J$ in the single digit percent regime as found by DFT+$U$ do not affect the phase digram qualitatively). We distinguish metallic, d-wave and spin density wave behavior. The false color plot indicates the estimated transition temperature. (d) Influence of the Hund's coupling $J$ on the transition temperature for different values of $U/t$. DFT+$U$ puts $J/U$ in the single digit percent regime, where its influence does not to alter the found phases, but can increase the transition temperature. (e) False color plot of the dispersion relation in the full Brillouin zone (non-shaded area). The paths plotted for the dispersion relation in Figs.~\ref{fig:fig1} and \ref{fig:fig2} are shown as dashed lines.  }
\label{fig:fig3}
\end{figure}

DFT+$U$ is routinely used to describe transition metal oxides. A study of TBBN is quite different, as the localization occurs at a much larger spatial scale, and the effective electronic parameters are expected to be much smaller.
Extracting quantitative parameters at finite doping is currently numerically not feasible and we considern just the undoped scenario.
Notwithstanding, in order to address the role of $J$ and $U$ at least in this simpler case, we use the recently proposed ACBN0 functional, which allows for a DFT-based consistent \textit{ab-initio} calculation of the value of $U$ and $J$ (see Methods section).
For each system, from  $7.34^\circ$ to $2.64^\circ$, we constructed the ``single shot'' Wannier states corresponding to the flat bands, and then computed the effective $U_{\mathrm{eff}}=U-J$.
As shown in Fig.~\ref{fig:fig3}(a), we obtained an $U_{\mathrm{eff}}$ that scales linearly with the twist angle, similar to what was proposed for TBG \cite{cao2018b}. The calculated scaling of $U_{\mathrm{eff}}$ is 34.9 meV/degree. Compared with the value of about 5 to 26 meV/degree proposed for TBG \cite{cao2018b}, this relatively large value suggests a quite strong electron-electron correlation effect in TBBN, probably due to the smaller dielectric constant in hBN.

In order to asses the role of $J$, we also computed the so-called Hubbard-Kanamori parameters $U$, $U'$, and $J$, using the bare Coulomb interaction (see Methods section). Our results yield two important results: i) we found that $(U-U')/2$ is always very close to $J$ (which is a relation strictly motivated only for the case of cubic symmetry for $d$ orbitals \cite{Werner17}). ii) the ratio of $J/U$ is found to be small (between 0.03 to 0.05 for the various angles).
These results are consistent with previous assumptions made for TBG, but are here confirmed using \textit{ab initio} simulations.
At finite doping, screening will decrease the values of the interaction parameters, which unfortunately is beyond an \textit{ab initio} analysis at present. We thus next, consider the effect of correlations, keeping $U$ and $J$ as free parameters, but assuming $J=(U-U')/2$.

We now turn to the construction of an effective low-energy model to deal with the inclusion of correlations, which can drive the system into different phases of matter. We find that the dispersion relation of the bands labeled by 3 and 7 in Fig.~\ref{fig:fig2} are fitted well by a two-band nearest neighbor hopping $t$ tight binding model on a triangular lattice (in agreement with the corresponding charge density plots in  Fig~\ref{fig:fig2}). In particular, this fit works well irrespective of the twist angle. This implies that the twist angle only appears as an indirect parameter of our model, via $U$, $J$, and the hopping $t$ (which is directly taken as $1/9$ of the DFT bandwidth). Therefore, analyzing the results in terms of $U/t$, $J/t$ and band-filling covers all the twist angles at once.

The role of Coulomb interaction versus phonons in TBBN is unclear at the moment. If phonons dominate we would expect an instability to emerge in the paring channel driving the system into a (BCS-type) s-wave superconducting phase, similar to what has been proposed for TBG \cite{lian18,Choi18,Wu18}. If on the other hand Coulomb repulsion is dominant, the fragile balance of competing orders gives rise to a zoo of different possible phases in dependence of interaction strength and filling \cite{Kennes2018d,Kang18,Yuan18,Koshino18,Ying18,Masayuki18,Chittari19}. To explore the later scenario, we employ a FRG approach (see Methods), which can be viewed as a renormalization group enhanced random phase approximation treatment of the interacting system. To this end we include an onsite Hubbard-type of repulsion $U$ as well as a weak Hund's coupling \cite{Xu18} term $J$ (in agreement with the above DFT+$U$ analysis). The dispersion relation in the full Brillouin zone is illustrated in a false color plot in Fig.~\ref{fig:fig3} (e). We define the chemical potential $\mu$ with respect to the van Hove filling (meaning that at $\mu=0$ the density of   states exhibits a van Hove singularity cf. Fig.~\ref{fig:fig2} (d) and (f)). At this filling the Fermi surface also exhibits nesting giving rise to a competition between a tendency for $d\pm id$ superconductivity  and a spin density wave as the interaction is turned on. This competition can be studied in an unbiased fashion using the FRG for small to intermediate $U/t$ by considering the so-called effective two-particle interaction $\Gamma$. During the renormalization group flow this effective two-particle interaction flows to strong coupling (diverges), which can be seen as the onset of ordering in the system. The energy scale at which this divergence occurs can be identified roughly with the critical temperature scale $T_c$ of the corresponding order. The momentum structure of the effective two-particle interaction indicates the type of ordering. We here parameterize the three independent momenta of the effective two-particle interaction $\Gamma(k_1,k_2,k_3)$ by their projection on the Fermi surface as well the angles $\phi_i$ of this projection. The calculated FRG phase diagram is summarized in Fig.~\ref{fig:fig3} (c) for zero Hund's coupling $J/t=0$. Turning on small Hund's couplings of up to 18\% changes $T_c$ as shown in Fig.~\ref{fig:fig3} (d), but does not influence the phase-diagram qualitatively, reinforcing the idea that the Hund's coupling $J$ does not play a major role here. Close to van Hove filling ($\mu=0$) we find topological $d\pm id$ superconductivity in this model for not too large $U/t$ whereas it turns into a spin-density wave (SDW) for larger values of $U/t$ \cite{Thomson18}. Further away from $\mu=0$ we also find metallic behavior which means that up to the point in energy space where the flow was stopped no divergence was encountered in the effective two-particle interaction. Fig.~\ref{fig:fig3} (b) illustrate the momentum structure of the effective two-particle interaction in the phase of $d\pm id$ superconductivity. The dominant diagonal features with sign change (blue to red) indicate that $d\pm id$ superconductivity is the dominant ordering tendency in doped TBBN.

Finally, we note that our \textit{ab initio} results show a linear dependence of $U$ in the twist angle, and a polynomial (order 2) variation of the bandwidth (and hence of $t$) in the twist angle for angle $>2^\circ$ (see Fig.~S3 in the supplemental materials). These two results implies that the ratio $U/t$ decreases as the twist angle increases and the system can be mechanically tuned through a transition from the SDW to the  d-wave superconducting phase by tuning the twist angle.

In sum, here we propose twisted bilayer boron nitride as a fascinating material in which a zoo of different electronic phenomena are expected. If the twist angle between the layers is small, groups of two-, three- and four-fold degenerate and energetically separated flat-bands show up. These do not rely crucially on how close the chosen angle is to a ``magic'' angle as in the recently studied twisted bilayer graphene.  Furthermore, at small angles the flat bands are truly separated from the valence and conduction bands, rendering twisted bilayer boron nitride  an ideal candidate to  study the physics of strong correlations in these materials. A combined DFT+$U$ and FRG ansatz reveals that the physics of the group of bands that are two-fold degenerate should be very similar to that of twisted bilayer graphene, possibly supporting topological d-wave superconductivity upon hole (or electron) doping. The existence of multiple flat bands significantly enriches the physics that can be probed. E.g. the separation of the flat bands found is in the 100meV regime. Coupling the present findings with the excitement of hyperbolic phonons \cite{Dai14} opens the possibility of having a strong phonon mediated coupling between the flat bands as the distance between them is in quasi-resonance with these phonon modes. Photodoping the system is another particularly intriguing avenue of future research. We predict that using photodoping the identified correlated phases (insulator and superconducting ones) can be induced in a transient, non-equilibrium manner, strongly motivating experimental advances along these lines. Our study also opens the door to similar theoretical work in other 2D materials.

\hspace{0.5in}

\noindent\emph{Acknowledgements} \\
Useful discussions with Martin Claassen and Mei-Yin Chou are acknowledged.
This work was supported by the European Research Council (ERC-2015-AdG694097) and Grupos Consolidados (IT578-13). The Flatiron Institute is a division of the Simons Foundation. LX acknowledges the European Unions Horizon 2020 research and innovation programme under the Marie Sklodowska-Curie grant agreement No. 709382 (MODHET).

\emph{Competing financial interests:}
The authors declare no competing financial interests.

\bibliography{TwistedBN}

\section*{Methods}

\paragraph*{Details about DFT Treatment}

The ground state DFT calculations are performed with the Vienna Ab initio simulation package (VASP) \cite{kresse93ab}. Plane waves are employed as a basis with an energy cutoff of 400 eV. The pseudo potentials are generated with the projector augmented wave method (PAW) \cite{blochl94} and the exchange-correlation potentials are treated in the local density approximation (LDA) \cite{perdew81}. As the system size is relatively large, a 1x1x1 kgrid is employed in the ground state and relaxation calculations. Lattice constants are chosen such that they correspond to a lattice constant of 2.512 Angstrom for a 1x1 unit cell of hBN. During the relaxation, all the atoms are relaxed until the force on each atom is less than 0.01 eV/Angstrom. The VESTA code \cite{vesta} is used for the visualization of the charge density distributions of the low-energy states of TBBN.

\paragraph*{DFT+$U$ calculations}
The \textit{ab initio} evaluation of the effective $U_{\mathrm{eff}}$ and the DFT+$U$ calculations are performed using the Octopus code \cite{Marques03,Castro06,Andrade15,Nicolas17}. A real-space spacing of 0.45 Bohr is chosen, and we employ norm-conserving pseudopotentials. The LDA is used for describing the local DFT part, and we utilize the ACBN0 functional to evaluate self consistently the effective $U_{\mathrm{eff}}$ of DFT+$U$. The vacuum size, atomic coordinates and lattice constant are taken to be the same as for the DFT treatment, as described above. Again, only the Gamma point is considered here. \\
The localized subspace is constructed as a single-shot Wannier states, taking the flat bands from LDA calculations. As the flat bands are energetically separated from the other occupied bands, this implies that the Wannier states reduce to the LDA states at the Gamma point.
\\
The Hubbard-Kanamori parameters $U$, $U'$ and $J$ are commonly defined from the expression
\begin{eqnarray}
 U = \frac{1}{N}\sum_{m} \langle mm | V_{ee} | mm\rangle\,,\nonumber\\
 J = \frac{1}{N(N-1)} \sum_{m\neq m'} \langle mm' | V_{ee} | m' m\rangle\,, \nonumber\\
 U' = \frac{1}{N(N-1)} \sum_{m\neq m'} \langle mm | V_{ee} | m' m'\rangle\, ,\nonumber
\end{eqnarray}
where $V_{ee}$ denotes the screened Coulomb interaction.

To evaluate these parameters, we extended the definition of the ACBN0 function which yields
\begin{eqnarray}
\bar{U} = \frac{\sum_{m}\sum_{\sigma}\bar{P}_{mm}^{\sigma}\bar{P}_{mm}^{-\sigma}(mm|mm)}{ \sum_{m}\sum_{\sigma} n_{mm}^{\sigma}n_{mm}^{-\sigma}}\,, \nonumber\\
\bar{U'} = \frac{\sum_{\{m\}}^*\sum_{\sigma\sigma'}\bar{P}_{mm'}^{\sigma}\bar{P}_{m'm''}^{\sigma'}(mm'|m''m''')}{\sum_{m\neq m'}\sum_{\sigma} [n_{mm}^{\sigma}n_{m'm'}^{\sigma} + n_{mm}^{\sigma}n_{m'm'}^{-\sigma}]}\,, \nonumber\\
\bar{J} = \frac{\sum_{\{m\}}^*\sum_{\sigma}\bar{P}_{mm'}^{\sigma}\bar{P}_{m'm''}^{\sigma}(mm'''|m''m')}{\sum_{m\neq m'}\sum_{\sigma} n_{mm}^{\sigma}n_{m'm'}^{\sigma} }\,,\nonumber
\end{eqnarray}
where $n_{mm'}^\sigma$ denotes the density matrix of the localized subspace, $\bar{P}_{mm'}^\sigma$ is the the renormalization density matrix (see Ref. \onlinecite{Nicolas17} for details), and $(mm'|m''m''')$ refers to the Coulomb integrals computed from the bare Coulomb interaction. In the sum over the orbitals $\sum_{\{m\}}^*$, the asterisk means that the sum goes over all the orbitals (two here) except for the case in which $m=m'=m''=m'''$. Importantly, we do not use any symmetry for computing the Coulomb integrals, and evaluate them directly on the real space grid, in order to analyze if the relation $\bar{U}-\bar{U'}=2\bar{J}$ is fulfilled or not.\\
We tested both the fully-localized-limit and the around-mean-field double counting terms, and found no sizable change in the value of $U_{\mathrm{eff}}$.

\paragraph*{Treating Correlations}

To treat correlations we introduce a tight binding model motivated from the DFT analysis. We want to concentrate on the bands labeled by 3 and 7 in Fig.~\ref{fig:fig1}. These can be approximated very well by a two band nearest neighbor tight binding model on a triangular lattice
\begin{equation}
H^0=-t\sum \limits_{\left\langle i,j \right\rangle}\sum\limits_{b=1}^2\sum_{\sigma=\uparrow,\downarrow} c_{i,b,\sigma}^\dagger c_{j,b,\sigma}-2t\sum \limits_{i}\sum\limits_{b=1}^2\sum_{\sigma=\uparrow,\downarrow} n_{i,b,\sigma},
\end{equation}
where $c_{i,b,\sigma}^{(\dagger)}$ annihilates (creates) a particle on site $i$ in the band $b$ and with spin $\sigma$ and $n_{i,b,\sigma}=c_{i,b,\sigma}^\dagger c_{i,b,\sigma}$. The second term (energy-shift) ensures that we measure the chemical potential with respect to van Hove filling for convenience.

We add local Coulomb repulsion and Hund's coupling to this by including
\begin{align}
H^U_i=&U\sum\limits_{b=1}^{2}n_{i,b,\uparrow}n_{i,b,\downarrow}+U'\sum\limits_{\sigma=\uparrow,\downarrow}n_{i,1,\sigma}n_{i,2,-\sigma}+(U'-J)\sum\limits_{\sigma=\uparrow,\downarrow}n_{i,1,\sigma}n_{i,2,\sigma}\notag\\
&-J'\left(c^\dagger_{i,1,\downarrow}c^\dagger_{i,2,\uparrow}c_{i,2,\downarrow}c_{i,1,\uparrow}+c^\dagger_{i,2,\uparrow}c^\dagger_{i,2,\downarrow}c_{i,1,\uparrow}c_{i,1,\downarrow}+{\rm H.c.}\right)
\end{align}
in the full Hamiltonian $H=H^0+\sum_i H^U_i$. To reduce the number of parameters we set $J=J'$ and $U'=U-2J$ (which strictly can be shown to hold for d-orbitals in free space \cite{Werner17}). From our DFT+$U$ estimates we find that this relation is also approximately fulfilled for twisted boron nitride although there seems to be no symmetry argument why this has to be the case. We choose $U$ and $J$ as the two independent parameters to vary in the following, which fixes $U'$.

The equilibrium phases of the Hamiltonian $H$ can be analyzed in the regime of small to intermediate interaction strength by the well-established functional renormalization group approach \cite{Metzner2012a}. Similar in spirit to other renormalization groups schemes within this method high-energy degrees of freedom are successively integrated out, resulting in a renormalized, effective low-energy theory of the problem. In our approach we monitor the behavior of the effective two-particle interaction over the flow, which addressed the successive inclusion of high-energy processes into an appropriate effective low energy model. Precursors to ordering tendencies are then indicated by a divergence of the effective two-particle interaction, signaling an instability of the Fermi-surface and a flow to strong coupling.

It is a particular advantage of the functional renormalization group that the method solely takes the Hamiltonian as an input and no a priori identification of
dominant interaction channel (e.g. magnetic, superconducting, ...) needs to be made by hand. This allows to treat these different channels on equal footing and thus analyze their competition in an unbiased fashion. As energy scales are successively integrated out over the flow, approaching the chemical potential $\mu$ the effective two-particle interaction acquires a strong momentum dependence. It is this momentum dependence combined with the analysis of the diverging channel that allows to draw conclusions about the dominant ordering in the system (varying the parameters of the Hamiltonian). Within the implementation of the FRG we use the implementation of Refs.~\cite{Metzner2012a,Kennes2018d} we start the renormalization group flow from the bare two-particle interaction of the Hamiltonian $H^U$, which shows no momentum dependence (owed to its locality). Therefore, initially ($\Lambda_i$ denoting the start of the flow) the effective two particle interaction $\Gamma^{\Lambda_i}(k_1,k_2,k_3,\omega_1,\omega_2,\omega_3)$, which in general depends on three independent momenta and frequencies, is momentum and frequency independent. The functional renormalization group approach then provides a recipe of how $\Gamma^\Lambda(k_1,k_2,k_3,\omega_1,\omega_2,\omega_3)$ at a different value of the flow parameter $\Lambda$ can be obtained along the line from $\Lambda_i$ all the way to $\Lambda_e$, which is the end of the flow where the effective low-energy theory of the initial model can be extracted. This is done in terms of rather cumbersome differential equation given in Ref.~\cite{Metzner2012a}.  When an ordering tendency is encountered during this successive flow the effective two-particle interaction $\Gamma^\Lambda(k_1,k_2,k_3,\omega_1,\omega_2,\omega_3)$ diverges at a finite $\Lambda_c$ and the flow needs to be stopped. This $\Lambda_c$ can roughly be identified with the critical temperature $T_c$ associated with the ordering tendencies.

During the flow (employing momentum and frequency conservation) the effective two-particle interaction depends on three independent momenta as well as three independent frequencies (it is associated with four Fermionic operators, like the bare two-particle interaction). To make the numerical calculation feasible we  have to reduce these degrees of freedoms. Since we are interested in the low-energy (equilibrium) physics of the model we project all frequencies to the chemical potential and all momenta to the Fermi-surface, but keeping their angle dependence, which we discretize in 18 uniform steps from $0$ to $2\pi$. For the equilibrium low-energy physics of a system, processes far away from the Fermi-surface are irrelevant, while the angle dependence of the on-shell processes are relevant; justifying this approximation. Therefore in our formulation of the functional renormalization group the low-energy effective two-particle interaction on the Fermi surface is parametrized by the three angles $\phi_1,\phi_2,\phi_3$ of momenta $k_1,k_2,k_3$ and thus it can be equivalently written as $\Gamma(\phi_1,\phi_2,\phi_3)$.

When a divergence is encountered in a particular ordering channel of the effective two-particle interaction,  with corresponding  order parameter $\hat\Delta_k$, one can employ a mean-field decomposition
\begin{equation}
\sum_{k,q}= \tilde\Gamma^{\Lambda_C}(k,q) [\hat\Delta_k,\hat\Delta_q]	
\end{equation}
of the dominant contribution to the vertex {\it{after}} the flow. The prefactor $\tilde\Gamma^{\Lambda_C}(k,q)$ can then be decomposed into the irreducible representation of the underlying lattice model to find the symmetry (e.g. s,p,d,...) of the dominant ordering.

\end{document}